\documentclass{IEEEtran}

\usepackage{amssymb,amsmath,bm}
\usepackage{graphicx,url}
\usepackage{cite}

\newlength\figwidth
\begin{document}

\title{On the Security of the Yi-Tan-Siew Chaotic Cipher%
\thanks{This paper has been published in \textit{IEEE Transactions on Circuits and Systems--II: Express Briefs}, vol. 51, no. 12, pp. 665-669, 2004.}
\thanks{This research was supported by the Applied R\&D Center, City
University of Hong Kong, under Grants 9410011 and 9620004.}}
\author{Shujun~Li, Guanrong~Chen, \IEEEmembership{Fellow,
IEEE}\thanks{Shujun Li and Guanrong Chen are with the Department
of Electronic Engineering, City University of Hong Kong, Kowloon,
Hong Kong, China.} and Xuanqin~Mou\thanks{Xuanqin Mou is with the
School of Electronics and Information Engineering, Xi'an Jiaotong
University, Xi'an, Shaanxi 710049, China.}}

%\pubid{0000--0000/00\$00.00~\copyright~2004 IEEE}

\maketitle

\begin{abstract}
This paper presents a comprehensive analysis on the security of
the Yi-Tan-Siew chaotic cipher proposed in
\cite{Yi:ChaoticCipher:IEEETCASI2002}. A differential
chosen-plaintext attack and a differential chosen-ciphertext
attack are suggested to break the sub-key $K$, under the
assumption that the time stamp can be altered by the attacker,
which is reasonable in such attacks. Also, some security Problems
about the sub-keys $\alpha$ and $\beta$ are clarified, from both
theoretical and experimental points of view. Further analysis
shows that the security of this cipher is independent of the use
of the chaotic tent map, once the sub-key $K$ is removed via the
proposed suggested differential chosen-plaintext attack.
\end{abstract}

\section{Introduction}

\PARstart{S}{ince} the 1990s, chaotic cryptography has attracted
more and more attention as a promising way to design novel
ciphers, and this research has become more intensive in recent
years\cite[Chap. 2]{ShujunLi:Dissertation2003}. To evaluate the
security performance of chaotic ciphers and to clarify some design
principles, cryptanalysis plays an important role.

This paper analyzes the security of the recently-proposed
Yi-Tan-Siew chaotic cipher \cite{Yi:ChaoticCipher:IEEETCASI2002}
and points out some defects existing in this cipher:
\begin{enumerate}
\item the sub-key $K$ can be removed by a differential
chosen-plaintext attack and a differential chosen-ciphertext
attack, under the assumption that the time-stamp $t$ can be
altered by the attacker;

\item the sub-key $\beta$ should not be contained in the secret
key due to its poor contribution to the security of the cipher;

\item the noise vectors $\{U_j\}$ used in the
encryption/decryption functions do not have a uniform
distribution, which downgrades the security of the cipher by
limiting the value of the sub-key $\alpha$;

\item when the aforementioned differential chosen-plaintext (or
chosen-ciphertext) attack is used, the security of the cipher is
independent of the chaotic map, but depends on the mixture of
three operations from different algebraic groups.
\end{enumerate}

The first two defects mean that the claimed key
$(\alpha,\beta,\gamma,K)$ collapses to be $(\alpha,\gamma)$. Note
that the second and third defects were implicitly mentioned in
Sec. III-B of \cite{Yi:ChaoticCipher:IEEETCASI2002} without
convincing explanations. This paper will give a comprehensive
analysis on all the four security defects.

The rest of this paper is organized as follows. The next section
gives a brief introduction to the Yi-Tan-Siew chaotic cipher.
Then, the first two defects of the cipher are discussed in Sec.
\ref{section:KeySpaceReduction}. The other two defects are
analyzed in Secs. \ref{section:NonUniformity} and
\ref{section:SecurityIndependentChaos}, respectively. The last
section concludes the paper.

\section{Yi-Tan-Siew Chaotic Cipher}

This proposed cipher is a \textit{time-variant} block cipher based
on the chaotic tent map. Each block has $4n$ bits, and the
encryption function changes as the iteration evolves. Given a
plaintext $P=(P_1,\cdots,P_j,\cdots,P_r)$ and the corresponding
ciphertext $C=(C_1\cdots,C_j,\cdots,C_r)$, where $P_j$ and $C_j$
are both $4n$-bit blocks, the cipher is described as follows.

\begin{itemize}
\item \textit{The employed chaotic tent map} is an extended
version of the normal skew tent map $F_\alpha$:
\begin{equation}
G_{(\alpha,\beta)}:x_i=
\begin{cases}
F_\alpha(x_{i-1}), & \mbox{if }0<x_{i-1}<1,\\
\beta, & \mbox{otherwise},
\end{cases}
\end{equation}
where
\begin{equation}
F_\alpha:x_i=
\begin{cases}
x_{i-1}/\alpha, & 0\leq x_{i-1}\leq\alpha,\\
(1-x_{i-1})/(1-\alpha), & \alpha<x_{i-1}\leq 1.
\end{cases}
\end{equation}

\item \textit{The secret key} was claimed to be a 4-tuple key
$(\alpha,\beta,\gamma,K)$, where $\gamma$ is used to generate a
secret initial condition $x_0$ of $G_{(\alpha,\beta)}$ as follows:
\begin{equation}
x_0=F_\gamma^{4n}\left(\frac{10^{\left\lfloor\log_{10}t\right\rfloor}}{t}\right).
\end{equation}
Here, $t$ representes the current time-stamp transmitted over a
public channel. Since $\gamma$ is only used to generate $x_0$, the
secret key can also be considered as $(\alpha,\beta,x_0,K)$. Based
on the secret key, the following secret functions are calculated
for the encryption/decryption procedures:
\begin{enumerate}
\item \textit{A sequence of $4n$-bit noise vectors}
$U_j=(u_{4jn},u_{4jn+1},\cdots,u_{4jn+4n-1})\;(j=0,1,2,\cdots,)$
are generated from the digital chaotic orbit\footnote{In this
paper, the term ``digital chaotic orbit" is used to denote the
orbit of a chaotic map realized in a digital computer \cite[Chap.
2.5]{ShujunLi:Dissertation2003}.} of the extended tent map
$G_{(\alpha,\beta)}$ with the following rule:
\begin{equation}
u_i=\begin{cases}
0, & 0\leq x_i\leq\alpha,\\
1, & \alpha<x_i\leq 1.
\end{cases}
\label{equation:NoiseVector}
\end{equation}

\item \textit{A sequence of secret permutations} $w_{ji}$
($j=0,1,2,\cdots$; $i=1,\cdots,n$) are generated from $U_j$ and
the sub-key $K$, as follows:
$V_j=(v_{j1},v_{j2},\cdots,v_{jn})=U_j\oplus K$, where each
$v_{ji}$ corresponds to a function $w_{ji}$ that represents a
permutation of four integers $\{1,2,3,4\}$ (following Table 1 of
\cite{Yi:ChaoticCipher:IEEETCASI2002}).

\item \textit{A sequence of secret bit-permutation functions}
$f_j=f_{jn}\circ\cdots\circ f_{j1}\;(j=0,1,2,\cdots)$ are
generated as follows:
\begin{eqnarray}
f_{ji}(X) & = & f_{ji}(M_1,M_2,M_3,M_4) \nonumber\\
& = & [w_{ji}(M_1,M_2,M_3,M_4)]\lll 1,
\end{eqnarray}
where $X=M_1\times 2^{3n}+M_2\times 2^{2n}+M_3\times 2^n+M_4$ and
``$\lll 1$" is the 1-bit circular left-shift operation.

\item \textit{Another sequence of permutation functions}
$f_j^{-1}=f_{j1}^{-1}\circ\cdots\circ f_{jn}^{-1}$ are generated
as follows:
\begin{eqnarray}
& f_{ji}^{-1}(X)=f_{ji}^{-1}(M_1,M_2,M_3,M_4)= & \nonumber\\
& [w_{ji}^{-1}(M_1,M_2,M_3,M_4)]\ggg 1, &
\end{eqnarray}
where $f_{ji}^{-1}$ is the inverse function of $f_{ji}$, i.e.,
$f_{ji}^{-1}\left(f_{ji}(X)\right)=X$, and ``$\ggg 1$" is the
1-bit circular right-shift operation.
\end{enumerate}

\item \textit{The initialization procedure}: $C_0=U_0, P_0=U_1$.

\item \textit{The encryption procedure}:
\begin{equation}
C_j=f_{j-1}\left(P_j\oplus\left(C_{j-1}\boxplus
U_{j+1}\right)\right)\oplus\left(P_{j-1}\boxplus U_{j+1}\right),
\label{equation:encryption}
\end{equation}
where $\oplus$ denotes XOR and $a\boxplus b:=(a+b)\bmod 2^{4n}$.

\item \textit{The decryption procedure}:
\begin{equation}
P_j=f_{j-1}^{-1}\left(C_j\oplus\left(P_{j-1}\boxplus
U_{j+1}\right)\right)\oplus\left(C_{j-1}\boxplus U_{j+1}\right).
\end{equation}
\end{itemize}

\section{Reduction of Key Space}
\label{section:KeySpaceReduction}

This section discusses the reduction of the key space of the
Yi-Tan-Siew cipher, i.e., its first two security defects.

\subsection{The Differential Chosen-Plaintext Attack for Reducing $K$}
\label{section:KReduction-DifferentialAttack}

To break the Yi-Tan-Siew cipher via a chosen-plaintext attack, the
attacker has to make $t$ fixed during the attack, i.e., to make
the sub-key $x_0$ and the noise vector sequence $\{U_j\}$ fixed.
This can be done by intentionally altering the local clock of the
encryption machine, which is generally available since the
attacker can access the encryption machine in chosen-plaintext
attacks \cite{Schneier:AppliedCryptography96}. If $t$ is generated
from a public time service, the attacker can simply altering the
time signal transmitted over the public channel to alter $t$. In
the following, therefore, assume that $t$ is fixed for all chosen
plaintexts.

Assume $\{P_1,\cdots,P_{j-1},P_j\}$ and
$\{P_1,\cdots,P_{j-1},P_j'\}$ are two plaintexts. The difference
of the ciphertexts is as follows:
\begin{eqnarray}
\Delta C_j=C_j\oplus C_j' & = &
f_{j-1}\left(P_j\oplus\left(C_{j-1}\boxplus
U_{j+1}\right)\right)\nonumber\\
& & {} \oplus f_{j-1}\left(P_j'\oplus\left(C_{j-1}\boxplus
U_{j+1}\right)\right).\label{equation:DeltaC}
\end{eqnarray}
Assume $CU_j=C_{j-1}\boxplus U_{j+1}$. Then, Eq.
(\ref{equation:DeltaC}) is reduced to
\begin{equation}
\Delta C_j=f_{j-1}\left(P_j\oplus CU_j\right)\oplus
f_{j-1}\left(P_j'\oplus CU_j\right).
\end{equation}

Now, consider such a question: \textit{what can one observe from
$\Delta C_j$, if $P_j$ and $P_j'$ have only one different bit?}
Assume that
\begin{eqnarray*}
P_j & = &
\left(p_{4jn},\cdots,p_{4jn+i},\cdots,p_{4jn+(4n-1)}\right),\\
P_j' & = &
\left(p_{4jn},\cdots,\overline{p_{4jn+i}},\cdots,p_{4jn+(4n-1)}\right),
\end{eqnarray*}
and $CU_j=\left(cu_0,\cdots,cu_{4n-1}\right)$. It is obvious that
$P_j\oplus CU_j$ and $P_j'\oplus CU_j$ also have only one
different bit at the same position $i$. Thus, further assuming
that
\[
P_j\oplus
CU_j=(p_{4jn}',\cdots,p_{4jn+i}',\cdots,p_{4jn+(4n-1)}'),
\]
\[
P_j'\oplus
CU_j=(p_{4jn}',\cdots,\overline{p_{4jn+i}'},\cdots,p_{4jn+(4n-1)}'),
\]
one has
\begin{IEEEeqnarray*}{RL}
\Delta C_j= & f_{j-1}(p_{4jn}',\cdots,p_{4jn+i}',\cdots,p_{4jn+(4n-1)}')\\
& \oplus
f_{j-1}(p_{4jn}',\cdots,\overline{p_{4jn+i}'},\cdots,p_{4jn+(4n-1)}').
\end{IEEEeqnarray*}
Considering $f_{j-1}$ is a bit-permutation function, one has
$f_{j-1}\left(P_j\oplus CU_j\right)=(p'_{4jn+I_0},\cdots,
p'_{4jn+I_l}=p_{4jn+i}',\cdots,p'_{4jn+I_{4n-1}})$, and
$f_{j-1}\left(P_j'\oplus CU_j\right)=(p'_{4jn+I_0},\cdots,
\overline{p'_{4jn+I_l}}=\overline{p_{4jn+i}'},\cdots,p'_{4jn+I_{4n-1}})$,
where $I_0\sim I_{4n-1}\in\{0,1,\cdots,4n-1\}$ denote the permuted
positions of the $4n$ bits $p_{4jn}'\sim p_{4jn+(4n-1)}'$. As a
result,
\begin{equation}
\Delta
C_j=(\overbrace{0,\cdots,0}^{4n-l},1,\overbrace{0,\cdots,0}^{l-1})=2^{l-1},
\end{equation}
which means that the $i$-th bit of $\Delta P=P_j\oplus P_j'$ is
permuted to the $l$-th bit of $\Delta C=C_j\oplus C_j'$ by
$f_{j-1}$.

From the above discussion, one can immediately conclude that given
the following $(4n+1)$ plaintexts containing $j$ plain-blocks, the
secret bit-permutation function $f_{j-1}$ can be exactly
reconstructed:
\begin{eqnarray*}
P^{(*)} & = & (P^*,\cdots,P^*,P^*),\\
P^{(1)} & = & (P^*,\cdots,P^*,P_1),\\
& \cdots &\\
P^{(l)} & = & (P^*,\cdots,P^*,P_l),\\
& \cdots &\\
P^{(4n)} & = & (P^*,\cdots,P^*,P_{4n}),
\end{eqnarray*}
where $P^*\oplus P_l=2^{l-1}$. To get all $r$ secret permutation
functions $f_0\sim f_{r-1}$ for the decryption of ciphertexts
whose sizes are not larger than $r$, the number of required
plaintexts is $(4n+1)\times r$.

Since the sub-key $K$ is used only to determine $\{f_j\}$
(together with $U_j$), the reconstruction of $f_0\sim f_{r-1}$
means the reduction of $K$ from the whole secret key
$(\alpha,\beta,\gamma,K)$.

Note that it is generally difficult to derive $V_j$ from $f_j$,
due to the strong mixing of $v_{ji}$ and the bit-shifting
operations. That is, it is generally difficult to derive $K$ from
$f_j$, even when $U_j$ is known to the attacker.

\subsection{The Differential Chosen-Ciphertext Attack for Reducing $K$}
\label{section:KReduction-DifferentialAttack2}

Due to the similarity of the encryption and decryption procedures,
the above differential chosen-plaintext attack can be easily
generalized to a differential chosen-ciphertext attack. Here, the
attacker can make $x_0$ fixed during the attack by altering $t$
transmitted over the public channel, which is possible since
generally the attacker has a full control of the public channel.
In the differential chosen-ciphertext attack, one can replace the
$(4n+1)\times r$ chosen plaintexts in the above differential
chosen-plaintext attack with $(4n+1)\times r$ chosen ciphertexts.
As a result, one can get all $r$ inverses permutation functions,
$f_0^{-1}\sim f_{r-1}^{-1}$, which is equivalent to the $r$
permutation functions, $f_0\sim f_{r-1}$.

\subsection{Reduction of $\beta$}
\label{section:betaReduction}

In Sec. III-B of \cite{Yi:ChaoticCipher:IEEETCASI2002}, it was
said that ``most likely, $\beta$ does not act in the encryption
and decryption processes", without any explanation. Here, we will
theoretically verify this claim.

In the extended tent map $G_{\alpha,\beta}$, $\beta$ will have to
make influence on the cipher only after $x=0$ or 1. However, the
possibility that $x=0$ or 1 is so tiny that the impact of $\beta$
on the encryption/decryption procedures is computationally
negligible from the Probabilistic point of view.

Without loss of generality, assume that the map $G_{\alpha,\beta}$
is realized in $n$-bit computing precision and that the digital
chaotic orbit distributes uniformly in the discretized space,
which is reasonable due to the uniform invariant density function
of the skew tent map\cite{Baranovsky:PLCM:IJBC95}. So, the
Probability that $x=0$ or 1 is $p=2/2^n=1/2^{n-1}$. As a result,
from the mathematical expectation of the geometric distribution
\cite{GeometricDistribution}, the average position of the first
occurrence of the above event ($x=0$ or 1) is $1/p=2^{n-1}$.

For single-precision floating-point arithmetic, $n=30$ (two sign
bits are excluded), averagely $2^{29}=512$M iterations are needed
to activate the influence of $\beta$ on the encryption/decryption
procedures. This means averagely $2^{29}/8=64$M leading bytes of
the ciphertext can be successfully decrypted without any knowledge
of $\beta$. Similarly, when the double-precision floating-point
arithmetic ($n=62$) is used, the condition will become much worse:
averagely $2^{61}/8=2$GG leading cipher-bytes can be decrypted
without knowing $\beta$.

Therefore, in most (if not all) cases, $\beta$ is not meaningful
in the key. In fact, it is just a trivial parameter (not part of
the secret key) to avoid the digital chaotic orbit of the normal
skew tent map $F_\alpha$ to fall into the fixed point $x=0$.

As a summary, under the above differential chosen-plaintext
attack, the original key $(\alpha,\beta,\gamma,K)$ collapses to be
$(\alpha,\gamma)$. When the differential chosen-plaintext attack
is impossible, the original key $(\alpha,\beta,\gamma,K)$
collapses to be $(\alpha,\gamma,K)$.

\section{Non-Uniformity of Noise Vector $U_j$}
\label{section:NonUniformity}

In the encryption procedure of the Yi-Tan-Siew cipher, the noise
vector $U_{j+1}$ is used to mask the plaintext $P_j$ together with
the previous plaintext $P_{j-1}$ and the previous ciphertext
$C_{j-1}$. To enhance the potential capability of resisting
statistics-based attacks\cite{Schneier:AppliedCryptography96}, it
is desirable that $U_j$ distributes uniformly in the discrete
space $\left\{0,\cdots,2^{4n}-1\right\}$. However, as mentioned in
Sec. III-B of \cite{Yi:ChaoticCipher:IEEETCASI2002}, $U_j$ does
not distributes uniformly when $\alpha$ is close to 0 or 1. As a
suggestion, $0.49<\alpha<0.5$ was suggested in
\cite{Yi:ChaoticCipher:IEEETCASI2002}. However, neither
theoretical nor experimental analysis is given in
\cite{Yi:ChaoticCipher:IEEETCASI2002} to support this claim.

In this section, we investigate the theory underlying the
non-uniformity of $U_j$ over $\left\{0,\cdots,2^{4n}-1\right\}$.
In addition, it is pointed out that the non-uniformity of $U_j$ is
also very significant when $\alpha=0.5$, which was not noticed in
\cite{Yi:ChaoticCipher:IEEETCASI2002}.

\subsection{Non-Uniformity of $U_j$ when $\alpha\neq 0.5$}
\label{section:NonUniformity1}

In this subsection, it is shown that when $\alpha\neq 0.5$, the
closer the $\alpha$ is to 0 or 1, the more severe the
non-uniformity of $U_j$ will become. Strictly speaking,
$\alpha\neq 0.5$ can never lead to a uniform distribution.

\begin{figure}%[!htbp]
\centering
\includegraphics[width=\figwidth]{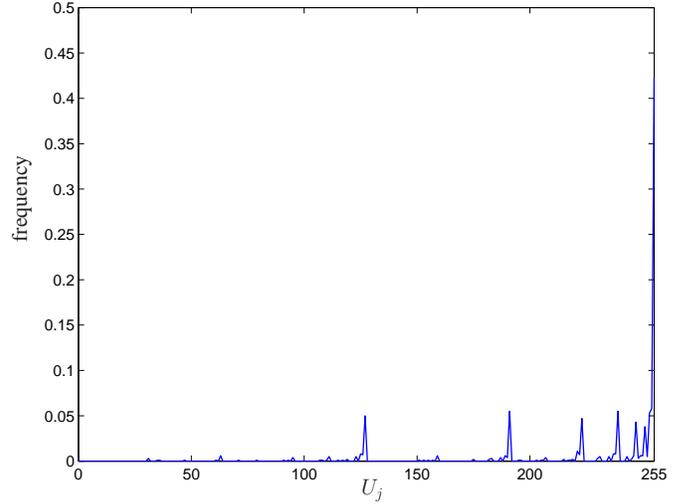}
\caption{The occurrence frequency of $U_j=a\in\{0,\cdots,255\}$,
when $(\alpha,\beta,x_0)=(0.1,0.7,0.3)$ (1000 samples).}
\label{figure:UjFrequency}
\end{figure}

Similar to Sec. \ref{section:betaReduction}, assume again that the
digital chaotic orbit of the map $G_{\alpha,\beta}$ distributes
uniformly in the discretized space. It is then easy to deduce the
following two Probabilities:
\begin{equation}
\mathrm{Prob}\{u_i=0\}=\alpha,\mathrm{Prob}\{u_i=1\}=1-\alpha.
\end{equation}
The above equations mean that $U_j$ will contain more 0-bits than
1-bits when $\alpha>0.5$, and more 1-bits than 0-bits when
$\alpha<0.5$. That is, $U_j$ does not have a uniform distribution
over $\left\{0,\cdots,2^{4n}-1\right\}$ if $\alpha\neq 0.5$. When
$(\alpha,\beta,x_0)=(0.1,0.7,0.3)$ and $n=2$, for example, under
double-precision floating-point arithmetic, Figure
\ref{figure:UjFrequency} gives an experimental curve of the
occurrence frequency of $U_j$ with different values between 0 and
$2^{4n}-1=2^8-1=255$. It can be seen that the frequency of
$U_j=255=(11111111)_2$ is close to 0.5 but many others are almost
0.

Under the assumption that all bits in $U_j$ are independent each
other, $\forall a\in\{0,\cdots,2^{4n}-1\}$, one can theoretically
deduce the Probability of $U_j=a$:
$\mathrm{Prob}\{U_j=a\}=\alpha^{N_0(a)}(1-\alpha)^{4n-N_0(a)}$,
where $N_0(a)\in\{0,\cdots,4n\}$ denotes the number of 0-bits in
$a$. In total there are $(4n+1)$ \textit{different} values in all
$2^{4n}$ Probabilities: $\mathrm{Prob}(0)=\alpha^{4n}$,
$\mathrm{Prob}(1)=\alpha^{4n}(1-\alpha)$, $\cdots$,
$\mathrm{Prob}(i)=\alpha^{4n-i}(1-\alpha)^i$, $\cdots$,
$\mathrm{Prob}(4n)=(1-\alpha)^{4n}$.

\begin{figure*}
\begin{eqnarray}
Com(\alpha) & = &
\sum\nolimits_{i=0}^{2n-1}\left(\mathrm{Prob}(i)\cdot\left(H(i)+\sum\nolimits_{m=0}^{\binom{4n}{i}}m\right)+
\mathrm{Prob}(4n-i)\cdot\left(H(i)+\binom{4n}{i}+\sum\nolimits_{m=0}^{\binom{4n}{4n-i}}m\right)\right)\nonumber\\
& & {}+\mathrm{Prob}(2n)\cdot\left(H(2n)+\sum\nolimits_{m=0}^{\binom{4n}{2n}}m\right)\nonumber\\
& = & {}
\sum\nolimits_{i=0}^{2n-1}\left(\left(\mathrm{Prob}(i)+\mathrm{Prob}(4n-i)\right)\cdot\left(H(i)+\frac{\binom{4n}{i}\left(\binom{4n}{i}+1\right)}{2}\right)+
\mathrm{Prob}(4n-i)\cdot\binom{4n}{i}\right)\\
& &
{}+\mathrm{Prob}(2n)\cdot\left(H(2n)+\frac{\binom{4n}{2n}\left(\binom{4n}{2n}+1\right)}{2}\right)\nonumber
\setcounter{equation}{13}\label{equation:QuickGuess}\setcounter{equation}{14}
\end{eqnarray}
\hrulefill%\vspace*{4pt}
\end{figure*}

The non-uniformity of each $U_j$ is useful for an attacker to get
its value more quickly via a specially-designed guess order. Since
the secret bit-permutation functions $\{f_{j-1}\}$ can be
reconstructed under chosen-plaintext attack (recall Sec.
\ref{section:KReduction-DifferentialAttack}), the attacker can
successfully decrypt any ciphertext once $\{U_j\}$ are obtained.
That is, $\left(\{U_j\},\{f_j\}\right)$ can be considered as an
equivalent of the original secret key $(\alpha,\beta,\gamma,K)$.

To find the right value of each $U_j$, the following guess order
of $U_j=a$ is suggested: $\forall a\in A_0\cup A_{4n}$, $\cdots$,
$\forall a\in A_i\cup A_{4n-i}$, $\cdots$, $\forall a\in A_{2n}$,
where $A_i$ ($i=0\sim 2n$) denotes the set of all $4n$-bit binary
integers that contain $i$ 0-bits. With such a guess order, the
average number of searched integers (i.e., the guess complexity)
$Com(\alpha)$ can be calculated with Eq.
(\ref{equation:QuickGuess}), where $H(i)$ denotes the number of
previous searched integers:
\begin{equation}
H(i)=\sum_{l=0}^{i-1}\binom{4n}{l}+\sum_{l=0}^{i-1}\binom{4n}{4n-l}=2\sum_{l=0}^{i-1}\binom{4n}{l}.
\end{equation}

\begin{figure}%[!htbp]
\centering
\includegraphics[width=\figwidth]{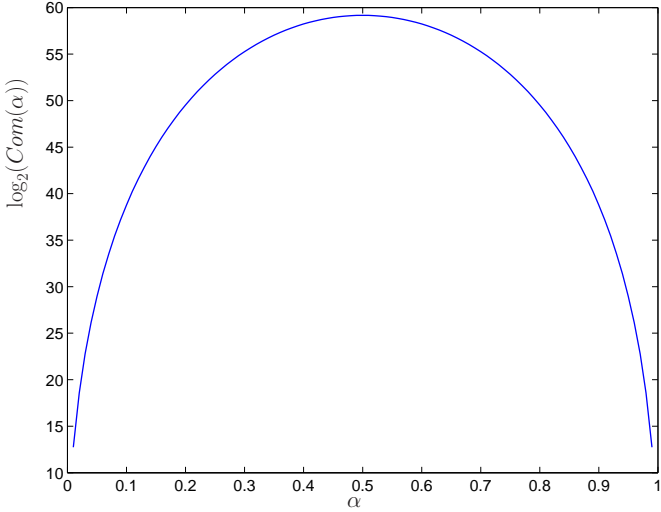}
\caption{$\log_2(Com(\alpha))$ vs.
$\alpha\in\{0.01,\cdots,0.01\times i,\cdots,0.99\}$.}
\label{figure:GuessComplexity}
\end{figure}

When $n=16$, for instance, the relationship between the calculated
complexity and the value of $\alpha$ is shown in Fig.
\ref{figure:GuessComplexity}. Note that there exist calculation
errors\footnote{The errors are natural results of the unavoidable
accumulation of the intermediate quantization errors.} that make
each $\log_2(Com(\alpha))$ a little less than the real value, but
this fact does not influence the following qualitative analysis.
From the experimental data given in Fig.
\ref{figure:GuessComplexity}, one can see that the complexity is
much less than $2^{4n-1}=2^{63}$ (the complexity of the
brute-force guess of a uniformly-distributed $4n$-bit integer)
when $\alpha$ is close to 0 or 1. Apparently, the closer the
$\alpha$ is to 0 or 1, the weaker the sub-key $\alpha$ will be. As
a result, to ensure the security of the Yi-Tan-Siew cipher, the
sub-key $\alpha$ has to be constrained in
$[\alpha_0,1-\alpha_0]\subset(0,1)$, where $Com(\alpha_0)$ should
be cryptographically large. This, however, will further reduce the
key space to some extent.

In \cite{Yi:ChaoticCipher:IEEETCASI2002}, $0.49<\alpha<0.5$ is
suggested to avoid this security defect. In this case, 1-bit will
always occur with a higher Probability than 0-bit, since
$\mathrm{Prob}\{u_i=1\}=1-\alpha>\mathrm{Prob}\{u_i=0\}=\alpha$.
So, one can guess the value of each $U_j$ with a different order:
$A_0\to\cdots\to A_{4n}$. Although
$\mathrm{Prob}\{u_i=1\}-\mathrm{Prob}\{u_i=0\}=1-2\alpha\in(0,0.02)$
is not so much, the guess complexity will still be less than the
simple brute-force search. From such a point of view,
$0.49<\alpha<0.5$ should be replaced by its balanced version:
$|\alpha-0.5|<0.01$.

\subsection{Non-Uniformity of $U_j$ when $\alpha=0.5$}
\label{section:NonUniformity2}

From the discussion given above, $\alpha=0.5$ seems to be the best
parameter to generate uniformly distributed $\{U_j\}$ that should
maximize the value of $Com(\alpha)$. Unfortunately, according to
our previous studies on the digital dynamics of piecewise-linear
chaotic maps (PWLCM) realized in fixed-point arithmetic
\cite[Chap. 3]{ShujunLi:Dissertation2003}, $\alpha=0.5$ is the
worst parameter from the viewpoint of dynamical degradation
occurring in the discretized space, which destroys the uniform
distribution of the generated pseudo-random numbers.

Actually, as a special case of the digital PWLCM, the digital
chaotic orbit of $G_{0.5,\beta}$ can be theoretically analyzed,
which is similar to but a little more complex than the orbit of
$F_{0.5}$. Without loss of generality, assume that the least
significant bit of $x_0$ is the $n_{x_0}$-th bit after the dot,
i.e., $x_0=(0.a_1a_2\cdots a_{n_{x_0}})_2$, where $a_{n_{x_0}}=1$.
To facilitate the following discussion, $n_{x_0}$ is called the
\textit{binary precision} of $x_0$. Substituting $\alpha=0.5$ into
the equation of $F_{0.5}$, one can get
\begin{equation}
F_{0.5}:x_i=
\begin{cases}
2\cdot x_{i-1}, & 0\leq x_{i-1}\leq 0.5,\\
2\cdot(1-x_{i-1}), & 0.5<x_{i-1}\leq 1.
\end{cases}
\end{equation}
It is obvious that $F_{0.5}(x_0)$ must be in the form of
$(0.a_1'a_2'\cdots a_{n_{x_0}-1}'0)_2$, where $a_{n_{x_0}-1}'=1$.
This means that the \textit{binary precision} of $x_0$ is
decreased by 1 after one iteration. Thus, the digital chaotic
orbit of $F_{0.5}$ will always trend to the same fixed point $x=0$
after $n_{x_0}$ iterations.

For $G_{0.5,\beta}$, the introduction of $\beta$ makes things a
little complicated: assuming that the \textit{binary precision} of
$\beta$ is $n_\beta$, the orbit of $G_{0.5,\beta}$ will be in the
following form:
\begin{eqnarray*}
& x_0\xrightarrow{n_{x_0}\mbox{
iterations}}0\to\beta\xrightarrow{n_\beta\mbox{
iterations}}0\to\beta\cdots &\\
& 0\to\beta\xrightarrow{n_\beta\mbox{ iterations}}0\to\beta\cdots
&
\end{eqnarray*}
That is, the digital chaotic orbit of $G_{0.5,\beta}$ enters a
periodic cycle determined by $\beta$ after a transient stage
determined by $x_0$. The period of the final cycle is $n_\beta+1$.

\begin{figure}%[!htbp]
\centering
\includegraphics[width=\figwidth]{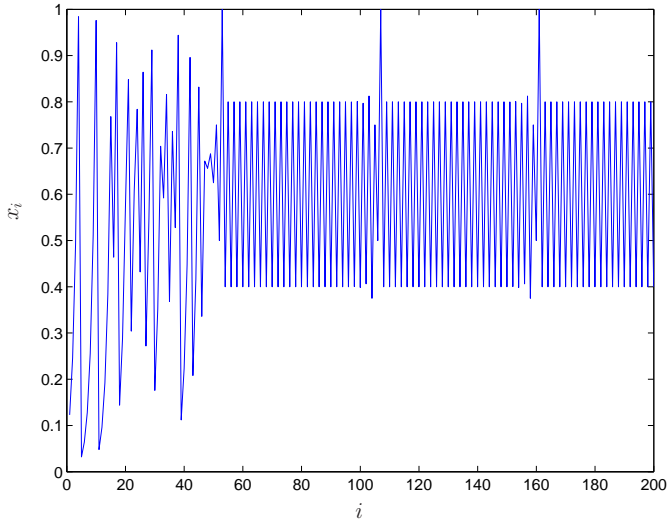}
\caption{The digital chaotic orbit of $G_{0.5,0.4}$ when
$x_0=0.123$.} \label{figure:TentHalf}
\end{figure}

As an example, when $x_0=0.123$, the digital chaotic orbit of
$G_{0.5,0.4}$ is shown in Fig. \ref{figure:TentHalf}. Apparently,
such a degraded chaotic orbit will generate badly non-uniform
$\{U_j\}$. When $n=2$, experiments show that the frequency of
$U_j=170$ is about $0.993$, which means that the non-uniformity is
even worse than the one given in Fig. \ref{figure:UjFrequency}.
One more example is also tested by changing the value of $\alpha$
in Fig. \ref{figure:UjFrequency} from 0.1 to 0.5 (but the values
of $\beta$ and $x_0$ are kept unchanged), and it is found that the
distribution of $U_j$ has two prominent peaks at $U_j=85$ and 170
(the frequencies are 0.412 and 0.418), respectively.

The above analysis shows that $\alpha=0.5$ is also a rather bad
parameter for the generation of $\{U_j\}$ toward a uniform
distribution. So, 0.5 should be excluded from the range of
$\alpha$. For example, the range $|\alpha-0.5|<0.01$ should be
replaced by $0<|\alpha-0.5|<0.01$.

\subsection{How to Mend This Defect?}

Since the non-uniformity of $\{U_j\}$ is mainly caused by the fact
that $\mathrm{Prob}\{u_i=0\}\neq \mathrm{Prob}\{u_i=1\}$, it is
easy to mend it by changing Eq. (\ref{equation:NoiseVector}) to
the following one:
\begin{equation}
u_i=\begin{cases}
0, & 0\leq x_i\leq 0.5,\\
1, & 0.5<x_i\leq 1.
\end{cases}
\end{equation}

It has been pointed out that dynamical degradation of
$G_{0.5,\beta}$ in the digital domain will influence the
uniformity of $\{U_j\}$. In fact, this Problem also exists for any
$\alpha\neq 0.5$, which has been clarified in \cite[Sec.
2.5.1]{ShujunLi:Dissertation2003}. Following previous studies, the
average length of all digital orbit of the tent map is
$O(2^{L/2})$, when $L$ is the bit number of the employed
finite-precision arithmetic. For double-finite floating-point
arithmetic, $L=62$, so the average length is about $2^{31}$, which
is not sufficiently large from the cryptographical point of view.
To overcome this Problem and also the non-uniformity caused by the
digital dynamical degradation, a small pseudo-random signal is
suggested to be used to perturb the digital chaotic orbit timely,
as discussed in Secs. 2.5.2 and 3.4.1 of
\cite{ShujunLi:Dissertation2003}.

\section{Incapability of Chaos for Security}
\label{section:SecurityIndependentChaos}

In Sec. \ref{section:NonUniformity1} above, it was mentioned that
$(\{U_j\},\{f_j\})$ is an equivalent of the original key. By
studying the possibility of solving for $\{U_j\}$ from chosen
plaintext-ciphertext pairs, it can be shown that the security of
the Yi-Tan-Siew cipher is independent of the use of the chaotic
map $G_{\alpha,\beta}$.

Given a plaintext $P_j$ and the corresponding ciphertext $C_j$,
one can get Eq. (\ref{equation:encryption}) for $U_{j+1}$. Under
the condition that $f_{j-1}$ has been reconstructed, it is
possible to solve for $U_{j+1}$ with a number of such equations.
Apparently, the solvability of $U_{j+1}$ is independent of the
chaotic map $G_{\alpha,\beta}$. That is, the security of the
cipher is independent of $G_{\alpha,\beta}$. In fact, one can
replace the chaotic map with any other PRNG to generate $U_j$,
without influencing the security of the cipher. Therefore, from
this point of view, the Yi-Tan-Siew cipher cannot be considered as
a typical chaotic cipher.

Next, the solvability of Eq. (\ref{equation:encryption}) is
discussed. Basically, the mixture of three different operations,
XOR, modulo $2^{4n}$ addition, and $f_{j-1}$, makes it rather
difficult to get $U_{j+1}$ from Eq. (\ref{equation:encryption}).
Rewrite Eq. (\ref{equation:encryption}) as follows:
\begin{equation}
C_j\oplus\left(P_{j-1}\boxplus
U_{j+1}\right)=f_{j-1}\left(P_j\oplus\left(C_{j-1}\boxplus
U_{j+1}\right)\right),
\end{equation}
which can be simplified as
\begin{equation}
a\oplus(b\boxplus x)=f_{j-1}(c\oplus(d\boxplus x)).
\end{equation}
The task is to find a $4n$-bit integer solution of $x$ from a
number of such equations. Considering that $f_{j-1}$ contains $n$
circular left-shift operations, it should have at least $2^n$
separate branches. This implies that at least $2^n$ points of
intersection between the graph of $a\oplus(b\boxplus x)$ and that
of $f_{j-1}(c\oplus(d\boxplus x))$ have to be checked to find the
only right integer solution of $x$. That is, a lower bound of the
complexity is $O(2^n)$.

\section{Conclusion}

This paper has studied the security of the recently-proposed
Yi-Tan-Siew chaotic cipher \cite{Yi:ChaoticCipher:IEEETCASI2002}.
Some defects of this cipher have been pointed out and analyzed in
detail. The security analyses given in this paper should provide
some useful references for better design of various chaotic
ciphers in the future.

\section*{Acknowledgements}

The authors would like to thank the anonymous reviewers for their
valuable comments.

\end{document}